\documentclass[aps,prl,showpacs,twocolumn,lineno,groupedaddress]{revtex4}  
\usepackage{graphicx}  
\usepackage{dcolumn}   
\usepackage{bm}        
\usepackage{amssymb}   

\begin{document}
%
%
\newcommand{\comm}[1]{\mbox{\mbox{\textup{#1}}}}
\newcommand{\subs}[1]{\mbox{\scriptstyle \mathit{#1}}}
\newcommand{\subss}[1]{\mbox{\scriptscriptstyle \mathit{#1}}}
\newcommand{\Frac}[2]{\mbox{\frac{\displaystyle{#1}}{\displaystyle{#2}}}}
\newcommand{\LS}[1]{\mbox{_{\scriptstyle \mathit{#1}}}}
\newcommand{\US}[1]{\mbox{^{\scriptstyle \mathit{#1}}}}
\def\gsim{\mathrel{\rlap{\raise.4ex\hbox{$>$}} {\lower.6ex\hbox{$\sim$}}}}
\def\lsim{\mathrel{\rlap{\raise.4ex\hbox{$<$}} {\lower.6ex\hbox{$\sim$}}}}
\renewcommand{\arraystretch}{1.3}
\newcommand{\Edep}{\mbox{E_{\mathit{dep}}}}
\newcommand{\Ebeam}{\mbox{E_{\mathit{beam}}}}
\newcommand{\Exrc}{\mbox{E_{\mathit{rec}}^{\mathit{exp}}}}
\newcommand{\Emrc}{\mbox{E_{\mathit{rec}}^{\mathit{sim}}}}
\newcommand{\Evis}{\mbox{E_{\mathit{vis}}}}
\newcommand{\Edepi}{\mbox{E_{\mathit{dep,i}}}}
\newcommand{\Evisi}{\mbox{E_{\mathit{vis,i}}}}
\newcommand{\Exrci}{\mbox{E_{\mathit{rec,i}}^{\mathit{exp}}}}
\newcommand{\Emrci}{\mbox{E_{\mathit{rec,i}}^{\mathit{sim}}}}
\newcommand{\Etmis}{\mbox{E_{\mathit{t,miss}}}}
%
%
\newcommand{\lt}{\mbox{$<$}}
\newcommand{\gt}{\mbox{$>$}}
\newcommand{\lte}{\mbox{$<=$}}
\newcommand{\gte}{\mbox{$>=$}}

\newcommand{\xa}{\mbox{$x_{a}$}}
\newcommand{\xb}{\mbox{$x_{b}$}}
\newcommand{\xp}{\mbox{$x_{p}$}}
\newcommand{\xpb}{\mbox{$x_{\bar{p}}$}}

\newcommand{\alphas}{\mbox{$\alpha_{s}$}}
\newcommand{\partt}{\mbox{$\partial^{t}$}}
\newcommand{\partd}{\mbox{$\partial_{t}$}}
\newcommand{\parmut}{\mbox{$\partial^{\mu}$}}
\newcommand{\parmud}{\mbox{$\partial_{\mu}$}}
\newcommand{\parnut}{\mbox{$\partial^{\nu}$}}
\newcommand{\parnud}{\mbox{$\partial_{\nu}$}}
\newcommand{\Amut}{\mbox{$A^{\mu}$}}
\newcommand{\Amud}{\mbox{$A_{\mu}$}}
\newcommand{\Anut}{\mbox{$A^{\nu}$}}
\newcommand{\Anud}{\mbox{$A_{\nu}$}}
\newcommand{\Fmunut}{\mbox{$F^{\mu\nu}$}}
\newcommand{\Fmunud}{\mbox{$F_{\mu\nu}$}}
\newcommand{\Gmut}{\mbox{$\gamma^{\mu}$}}
\newcommand{\Gmud}{\mbox{$\gamma_{\mu}$}}
\newcommand{\Gnut}{\mbox{$\gamma^{\nu}$}}
\newcommand{\Gnud}{\mbox{$\gamma_{\nu}$}}
\newcommand{\pvecsq}{\mbox{$\vec{p}^{\:2}$}}
\newcommand{\bigsum}{\mbox{$\displaystyle{\sum}$}}
%
\newcommand{\Lzero}{Level \O}
\newcommand{\Lone}{Level $1$}
\newcommand{\Ltwo}{Level $2$}
\newcommand{\Lhalf}{Level $1.5$}
%
\newcommand{\bzero}{\mbox{\comm{B\O}}}
\newcommand{\dzero}{\mbox{D\O}}
\newcommand{\dzerosm}{\mbox{\comm{$\scriptsize{D\O}$}}}
\newcommand{\runb}{Run~$1$B}
\newcommand{\runa}{Run~$1$A}
\newcommand{\runone}{Run~$1$}
\newcommand{\runtwo}{Run~$2$}
\newcommand{\dzpjet}{\textsc{D{\O}Pjet}}
\newcommand{\y}{\mbox{$y$}}
\newcommand{\z}{\mbox{$z$}}
\newcommand{\px}{\mbox{$p_{x}$}}
\newcommand{\py}{\mbox{$p_{y}$}}
\newcommand{\pz}{\mbox{$p_{z}$}}
\newcommand{\ex}{\mbox{$E_{x}$}}
\newcommand{\ey}{\mbox{$E_{y}$}}
\newcommand{\ez}{\mbox{$E_{z}$}}
\newcommand{\et}{\mbox{$E_{T}$}}
\newcommand{\etprime}{\mbox{$E_{T}^{\prime}$}}
\newcommand{\etone}{\mbox{$E_{T}^{\mathrm{1}}$}}
\newcommand{\ettwo}{\mbox{$E_{T}^{\mathrm{2}}$}}
\newcommand{\etlj}{\mbox{$E_{T}^{\subs{lj}}$}}
\newcommand{\etmax}{\mbox{$E_{T}^{max}$}}
\newcommand{\etcand}{\mbox{$E_{T}^{\subs{cand}}$}}
\newcommand{\etup}{\mbox{$E_{T}^{\subs{up}}$}}
\newcommand{\etdown}{\mbox{$E_{T}^{\subs{down}}$}}
\newcommand{\jet}{\mbox{$E_{T}^{\subs{jet}}$}}
\newcommand{\cet}{\mbox{$E_{T}^{\subs{cell}}$}}
\newcommand{\jetvec}{\mbox{$\vec{E}_{T}^{\subs{jet}}$}}
\newcommand{\cetvec}{\mbox{$\vec{E}_{T}^{\subs{cell}}$}}
\newcommand{\jevec}{\mbox{$\vec{E}^{\subs{jet}}$}}
\newcommand{\cevec}{\mbox{$\vec{E}^{\subs{cell}}$}}
\newcommand{\etfr}{\mbox{$f_{E_{T}}$}}
\newcommand{\aveet}{\mbox{$\langle\et\rangle$}}
\newcommand{\nj}{\mbox{$n_{j}$}}
\newcommand{\ptrel}{\mbox{$p_{T}^{rel}$}}
\newcommand{\etad}{\mbox{$\eta_{d}$}}           
\newcommand{\peta}{\mbox{$\eta$}}               
\newcommand{\aeta}{\mbox{$|\eta|$}}             
\newcommand{\ipb}{pb$^{-1}$}
\newcommand{\inb}{nb$^{-1}$}
\newcommand{\met}{\mbox{${\hbox{$E$\kern-0.63em\lower-.18ex\hbox{/}}}_{T}$}}
\newcommand{\metvec}{\mbox{${\hbox{$\vec{E}$\kern-0.63em\lower-.18ex\hbox{/}}}_{T}\,$}}
\newcommand{\metx}{\mbox{${\hbox{$E$\kern-0.63em\lower-.18ex\hbox{/}}}_{x}\,$}}
\newcommand{\mety}{\mbox{${\hbox{$E$\kern-0.63em\lower-.18ex\hbox{/}}}_{y}\,$}}
\newcommand{\het}{\mbox{$\vec{\mathcal{H}}_{T}$}}
\newcommand{\hetsc}{\mbox{$\mathcal{H}_{T}$}}
\newcommand{\zvrt}{\mbox{$Z$}}
\newcommand{\zcut}{\mbox{$|\zvrt| < 50$}}
\newcommand{\mtwo}{\mbox{$\mathcal{M}_{2}$}}
\newcommand{\mthree}{\mbox{$\mathcal{M}_{3}$}}
\newcommand{\mfour}{\mbox{$\mathcal{M}_{4}$}}
\newcommand{\msix}{\mbox{$\mathcal{M}_{6}$}}
\newcommand{\mn}{\mbox{$\mathcal{M}_{n}$}}
\newcommand{\R}{\mbox{$R_{\subss{MTE}}$}}
\newcommand{\invR}{\mbox{$1/R_{\subss{MTE}}$}}
\newcommand{\eemf}{\mbox{$\varepsilon_{\subss{EMF}}$}}
\newcommand{\echf}{\mbox{$\varepsilon_{\subss{CHF}}$}}
\newcommand{\ehcf}{\mbox{$\varepsilon_{\subss{HCF}}$}}
\newcommand{\eglob}{$\mbox{\varepsilon_{\subs{glob}}$}}
\newcommand{\emte}{\mbox{$\varepsilon_{\subss{MTE}}$}}
\newcommand{\ezvrt}{\mbox{$\varepsilon_{\subss{Z}}$}}
\newcommand{\etot}{\mbox{$\varepsilon_{\subs{tot}}$}}
\newcommand{\Bprime}{\mbox{$\comm{B}^{\prime}$}}
\newcommand{\Nsurv}{\mbox{$N_{\subs{surv}}$}}
\newcommand{\Nfail}{\mbox{$N_{\subs{fail}}$}}
\newcommand{\Ntot}{\mbox{$N_{\subs{tot}}$}}
\newcommand{\p}[1]{\mbox{$p_{#1}$}}
\newcommand{\ep}[1]{\mbox{$\Delta p_{#1}$}}
\newcommand{\delr}{\mbox{$\Delta R$}}
\newcommand{\deleta}{\mbox{$\Delta\eta$}}
\newcommand{\cafone}{{\sc Cafix 5.1}}
\newcommand{\caftwo}{{\sc Cafix 5.2}}
\newcommand{\delphi}{\mbox{$\Delta\varphi$}}
\newcommand{\rphi}{\mbox{$r-\varphi$}}
\newcommand{\etaphi}{\mbox{$\eta-\varphi$}}
\newcommand{\etatphi}{\mbox{$\eta\times\varphi$}}
\newcommand{\Rjet}{\mbox{$R_{jet}$}}
\newcommand{\jphi}{\mbox{$\varphi_{\subs{jet}}$}}
\newcommand{\gphi}{\mbox{$\varphi_{\subs{\gamma}}$}}
\newcommand{\ceta}{\mbox{$\eta^{\subs{cell}}$}}
\newcommand{\cphi}{\mbox{$\phi^{\subs{cell}}$}}
\newcommand{\inlum}{\mbox{$\mathcal{L}$}}
\newcommand{\gm}{\mbox{$\gamma$}}
\newcommand{\Rjj}{\mbox{$\mathbf{R}_{\subs{jj}}$}}
\newcommand{\Rgj}{\mbox{$\mathbf{R}_{\subs{\gamma j}}$}}
\newcommand{\Rmathcal}{\mbox{$\mathcal{R}$}}
\newcommand{\etv}{\mbox{$\vec{E}_{T}$}}
\newcommand{\nvec}{\mbox{$\hat{\vec{n}}$}}
\newcommand{\eprime}{\mbox{$E^{\prime}$}}
\newcommand{\aveprime}{\mbox{$\bar{E}^{\prime}$}}
\newcommand{\geta}{\mbox{$\eta_{\gm}$}}
\newcommand{\jeta}{\mbox{$\eta_{\subs{jet}}$}}
\newcommand{\cjeta}{\mbox{$\eta_{\subs{jet}}^{\subss{CEN}}$}}
\newcommand{\fjeta}{\mbox{$\eta_{\subs{jet}}^{\subss{FOR}}$}}
\newcommand{\cjphi}{\mbox{$\varphi_{\subs{jet}}^{\subss{CEN}}$}}
\newcommand{\fjphi}{\mbox{$\varphi_{\subs{jet}}^{\subss{FOR}}$}}
\newcommand{\etcut}{\mbox{$E_{T}^{\subs{cut}}$}}
\newcommand{\etg}{\mbox{$E_{T}^{\gm}$}}
\newcommand{\cenet}{\mbox{$E_{T}^{\subss{CEN}}$}}
\newcommand{\foret}{\mbox{$E_{T}^{\subss{FOR}}$}}
\newcommand{\cenen}{\mbox{$E^{\subss{CEN}}$}}
\newcommand{\foren}{\mbox{$E^{\subss{FOR}}$}}
\newcommand{\ejtptc}{\mbox{$E^{\subs{ptcl}}_{\subs{jet}}$}}
\newcommand{\ejtmes}{\mbox{$E^{\subs{meas}}_{\subs{jet}}$}}
\newcommand{\AIDA}{{\sc AIDA}}
\newcommand{\RECO}{{\sc Reco}}
\newcommand{\PYTHIA}{{\sc Pythia}}
\newcommand{\HERWIG}{{\sc Herwig}}
\newcommand{\JETRAD}{{\sc Jetrad}}
\newcommand{\CTone}{\mbox{$|\eta|<0.4}$}
\newcommand{\CTtwo}{\mbox{$0.4\leq|\eta|<0.8$}}
\newcommand{\ICone}{\mbox{$0.8\leq|\eta|<1.2$}}
\newcommand{\ICtwo}{\mbox{$1.2\leq|\eta|<1.6$}}
\newcommand{\FWone}{\mbox{$1.6\leq|\eta|<2.0$}}
\newcommand{\FWtwo}{\mbox{$2.0\leq|\eta|<2.5$}}
\newcommand{\FWthr}{\mbox{$2.5\leq|\eta|<3.0$}}
\newcommand{\LCTone}{\mbox{$|\eta|<0.5$}}
\newcommand{\LCTtwo}{\mbox{$0.5\leq|\eta|<1.0$}}
\newcommand{\LICone}{\mbox{$1.0\leq|\eta|<1.5$}}
\newcommand{\LICtwo}{\mbox{$1.5\leq|\eta|<2.0$}}
\newcommand{\LFWone}{\mbox{$2.0\leq|\eta|<3.0$}}
\newcommand{\CSone}{\mbox{$|\eta|<0.5$}}
\newcommand{\CStwo}{\mbox{$0.5\leq|\eta|<1.0$}}
\newcommand{\CSthr}{\mbox{$1.0\leq|\eta|<1.5$}}
\newcommand{\CSfou}{\mbox{$1.5\leq|\eta|<2.0$}}
\newcommand{\CSfiv}{\mbox{$2.0\leq|\eta|<3.0$}}
\newcommand{\sigA}{\mbox{$\sigma_{\subss{\!A}}$}}
\newcommand{\sigASS}{\mbox{$\sigma_{\subss{A}}^{\subss{SS}}$}}
\newcommand{\sigAOS}{\mbox{$\sigma_{\subss{A}}^{\subss{OS}}$}}
\newcommand{\sigZ}{\mbox{$\sigma_{\subss{Z}}$}}
\newcommand{\sige}{\mbox{$\sigma_{\subss{E}}$}}
\newcommand{\siget}{\mbox{$\sigma_{\subss{\et}}$}}
\newcommand{\sigetone}{\mbox{$\sigma_{\subs{\etone}}$}}
\newcommand{\sigettwo}{\mbox{$\sigma_{\subs{\ettwo}}$}}
\newcommand{\rcal}{\mbox{$R_{\subs{cal}}$}}
\newcommand{\zcal}{\mbox{$Z_{\subs{cal}}$}}
\newcommand{\Runf}{\mbox{$R_{\subs{unf}}$}}
\newcommand{\Rsep}{\mbox{$\mathcal{R}_{sep}$}}
\newcommand{\etal}{{\it et al.}}
\newcommand{\ppbar}{\mbox{$p\overline{p}$}}
\newcommand{\pp}{\mbox{$pp$}}
\newcommand{\qqbar}{\mbox{$q\overline{q}$}}
\newcommand{\ccbar}{\mbox{$c\overline{c}$}}
\newcommand{\bbbar}{\mbox{$b\overline{b}$}}
\newcommand{\ttbar}{\mbox{$t\overline{t}$}}

\newcommand{\bbj}{\mbox{$b\overline{b}j$}}
\newcommand{\bbjj}{\mbox{$b\overline{b}jj$}}
\newcommand{\ccjj}{\mbox{$c\overline{c}jj$}}

\newcommand{\bb}{\mbox{$b\overline{b}j(j)$}}
\newcommand{\cc}{\mbox{$c\overline{c}j(j)$}}

\newcommand{\hboson}{\mbox{$\mathit{h}$}}
\newcommand{\Hboson}{\mbox{$\mathit{H}$}}
\newcommand{\Aboson}{\mbox{$\mathit{A}$}}
\newcommand{\zboson}{\mbox{$\mathit{Z}$}}
\newcommand{\zb}{\mbox{$\mathit{Zb}$}}
\newcommand{\bh}{\mbox{$\mathit{bh}$}}
\newcommand{\btag}{\mbox{$\mathit{b}$}}

\newcommand{\hsm}{\mbox{$h_{SM}$}}
\newcommand{\hmssm}{\mbox{$h_{MSSM}$}}

\newcommand{\prot}{\mbox{$p$}}
\newcommand{\pbar}{\mbox{$\overline{p}$}}
\newcommand{\pt}{\mbox{$p_{T}$}}
\newcommand{\xnot}{\mbox{$X_{0}$}}
\newcommand{\Znot}{\mbox{$Z^{0}$}}
\newcommand{\Wpm}{\mbox{$W^{\pm}$}}
\newcommand{\Wplus}{\mbox{$W^{+}$}}
\newcommand{\Wminus}{\mbox{$W^{-}$}}
\newcommand{\lamb}{\mbox{$\lambda$}}
\newcommand{\nhatbf}{\mbox{$\hat{\mathbf{n}}$}}
\newcommand{\pbf}{\mbox{$\mathbf{p}$}}
\newcommand{\xbf}{\mbox{$\mathbf{x}$}}
\newcommand{\jbf}{\mbox{$\mathbf{j}$}}
\newcommand{\Ebf}{\mbox{$\mathbf{E}$}}
\newcommand{\Bbf}{\mbox{$\mathbf{B}$}}
\newcommand{\Abf}{\mbox{$\mathbf{A}$}}
\newcommand{\Rbf}{\mbox{$\mathbf{R}$}}
\newcommand{\nablabf}{\mbox{$\mathbf{\nabla}$}}
\newcommand{\rarrow}{\mbox{$\rightarrow$}}
\newcommand{\slashp}{\mbox{$\not \! p \,$}}
\newcommand{\slashk}{\mbox{$\not \! k$}}
\newcommand{\slasha}{\mbox{$\not \! a$}}
\newcommand{\slashA}{\mbox{$\! \not \! \! A$}}
\newcommand{\slashpar}{\mbox{$\! \not \! \partial$}}
\newcommand{\intdouble}{\mbox{$\int\!\!\int$}}
\newcommand{\MRSTGU}{MRSTg$\uparrow$}
\newcommand{\MRSTGD}{MRSTg$\downarrow$}
%
\newcommand{\Due}{\mbox{$D_{\mathrm{ue}}$}}
\newcommand{\Dth}{\mbox{$D_{\Theta}$}}
\newcommand{\Dof}{\mbox{$D_{\mathrm{O}}$}}
\newcommand{\zbl}{\texttt{ZERO BIAS}}
\newcommand{\mbl}{\texttt{MIN BIAS}}
\newcommand{\mbll}{\texttt{MINIMUM BIAS}}
\newcommand{\nue}{\mbox{$\nu_{e}$}}
\newcommand{\num}{\mbox{$\nu_{\mu}$}}
\newcommand{\nut}{\mbox{$\nu_{\tau}$}}
\newcommand{\mycs}{\mbox{$d^{\,2}\sigma/(d\et d\eta)$}}
\newcommand{\mycsav}{\mbox{$\langle \mycs \rangle$}}
\newcommand{\tdcs}{\mbox{$d^{\,3}\sigma/d\et d\eta_{1} d\eta_{2}$}}
\newcommand{\tdcsav}{\mbox{$\langle d^{\,3}\sigma/d\et d\eta_{1} d\eta_{2} \rangle$}}
\newcommand{\tanb}{$\tan\beta$}
\newcommand{\cotb}{$\cot\beta$}

\newcommand{\rstev}{\mbox{$\rs = \T{1.8}$}}
\newcommand{\rssps}{\mbox{$\rs = \T{0.63}$}}
\newcommand{\XX}{\mbox{$\, \times \,$}}
\newcommand{\AP}{\mbox{${\rm \bar{p}}$}}
\newcommand{\SU}{\mbox{$<\! |S|^2 \!>$}}
\newcommand{\ET}{\mbox{$E_{T}$}}
\newcommand{\HT}{\mbox{$S_{{\rm {\sl T}}}$} }
\newcommand{\PT}{\mbox{$p_{t}$}}
\newcommand{\DP}{\mbox{$\Delta\phi$}}
\newcommand{\DR}{\mbox{$\Delta R$}}
\newcommand{\DE}{\mbox{$\Delta\eta$}}
\newcommand{\DEP}{\mbox{$\Delta\eta_{c}$}}
\newcommand{\PH}{\mbox{$\phi$}}
\newcommand{\EA}{\mbox{$\eta$} }
\newcommand{\EAJ}{\mbox{\EA(jet)}}
\newcommand{\AEA}{\mbox{$|\eta|$}}
\newcommand{\Ge}[1]{\mbox{#1 GeV}}
\newcommand{\T}[1]{\mbox{#1 TeV}}
\newcommand{\x}{\cdot}
\newcommand{\ra}{\rightarrow}
\def\D0{D\O}
\def\ETmiss{{\rm {\mbox{$E\kern-0.57em\raise0.19ex\hbox{/}_{T}$}}}}
\newcommand{\mb}{\mbox{mb}}
\newcommand{\nb}{\mbox{nb}}
\newcommand{\rs}{\mbox{$\sqrt{\rm {\sl s}}$}}
\newcommand{\fdel}{\mbox{$f(\DEP)$}}
\newcommand{\fdele}{\mbox{$f(\DEP)^{exp}$}}
\newcommand{\fgap}{\mbox{$f(\DEP\! \geq \!3)$}}
\newcommand{\fgape}{\mbox{$f(\DEP\! \geq \!3)^{exp}$}}
\newcommand{\fpyt}{\mbox{$f(\DEP\!>\!2)$}}
\newcommand{\delth}{\mbox{$\DEP\! \geq \!3$}}
\newcommand{\uplim}{\mbox{$1.1\!\times\!10^{-2}$}}
\def\simge
{\mathrel{\rlap{\raise 0.53ex \hbox{$>$}}{\lower 0.53ex \hbox{$\sim$}}}}
\def\simle
{\mathrel{\rlap{\raise 0.53ex \hbox{$<$}}{\lower 0.53ex \hbox{$\sim$}}}}
\newcommand{\pbarp}{\mbox{$p\bar{p}$}}
\def\ETmiss{\mbox{${\hbox{$E$\kern-0.5em\lower-.1ex\hbox{/}\kern+0.15em}}_{\rm T}$}}
\def\Et{\mbox{$E_{T}$}}
\newcommand{\modeta}{\mid \!\! \eta \!\! \mid}
\def\gevcc{GeV/c$^2$}                   
\def\gevc{GeV/c}                        
\def\gev{GeV}                           
\newcommand{\als}{\mbox{${\alpha_{{\rm s}}}$}}
\def\1960{$\sqrt{s}=1960$ GeV}
\def\etI{E_{T_1}}
\def\etII{E_{T_2}}
\def\itaI{\eta_1}
\def\itaII{\eta_2}
\def\deta{\Delta\eta}
\def\etab{\bar{\eta}}
\def\xq{($x_1$,$x_2$,$Q^2$)}
\def\xx{($x_1$,$x_2$)}
\def\rap{pseudorapidity}
\def\as{\alpha_s}
\def\ap{\alpha_{\rm BFKL}}
\def\apb{\alpha_{{\rm BFKL}_{bin}}}
\def\cm{c.m.}


\hspace{5.2in}\mbox{FERMILAB-PUB-05/058-E}

\title{Search for neutral supersymmetric Higgs bosons in multijet events at $\sqrt{s}=$1.96~TeV}
%
\author{                                                                      
V.M.~Abazov,$^{35}$                                                           
B.~Abbott,$^{72}$                                                             
M.~Abolins,$^{63}$                                                            
B.S.~Acharya,$^{29}$                                                          
M.~Adams,$^{50}$                                                              
T.~Adams,$^{48}$                                                              
M.~Agelou,$^{18}$                                                             
J.-L.~Agram,$^{19}$                                                           
S.H.~Ahn,$^{31}$                                                              
M.~Ahsan,$^{57}$
% ...
\\                                                                            
\vskip 0.30cm                                                                 
\centerline{(D\O\ Collaboration)}                                             
\vskip 0.30cm                                                                 
}                                                                             
\affiliation{                                                                 
\centerline{$^{1}$Universidad de Buenos Aires, Buenos Aires, Argentina}       
\centerline{$^{2}$LAFEX, Centro Brasileiro de Pesquisas F{\'\i}sicas,         
                  Rio de Janeiro, Brazil}                                     
\centerline{$^{3}$Universidade do Estado do Rio de Janeiro,                   
                  Rio de Janeiro, Brazil}                                     
\centerline{$^{4}$Instituto de F\'{\i}sica Te\'orica, Universidade            
                  Estadual Paulista, S\~ao Paulo, Brazil}                     
\centerline{$^{5}$University of Alberta, Edmonton, Alberta, Canada,           
               Simon Fraser University, Burnaby, British Columbia, Canada,}   
\centerline{York University, Toronto, Ontario, Canada, and                    
         McGill University, Montreal, Quebec, Canada}                         
\centerline{$^{6}$Institute of High Energy Physics, Beijing,                  
                  People's Republic of China}                                 
\centerline{$^{7}$University of Science and Technology of China, Hefei,       
                  People's Republic of China}                                 
\centerline{$^{8}$Universidad de los Andes, Bogot\'{a}, Colombia}             
\centerline{$^{9}$Center for Particle Physics, Charles University,            
                  Prague, Czech Republic}                                     
\centerline{$^{10}$Czech Technical University, Prague, Czech Republic}        
\centerline{$^{11}$Institute of Physics, Academy of Sciences, Center          
                  for Particle Physics, Prague, Czech Republic}               
\centerline{$^{12}$Universidad San Francisco de Quito, Quito, Ecuador}        
\centerline{$^{13}$Laboratoire de Physique Corpusculaire, IN2P3-CNRS,         
                 Universit\'e Blaise Pascal, Clermont-Ferrand, France}        
\centerline{$^{14}$Laboratoire de Physique Subatomique et de Cosmologie,      
                  IN2P3-CNRS, Universite de Grenoble 1, Grenoble, France}     
\centerline{$^{15}$CPPM, IN2P3-CNRS, Universit\'e de la M\'editerran\'ee,     
                  Marseille, France}                                          
\centerline{$^{16}$Laboratoire de l'Acc\'el\'erateur Lin\'eaire,              
                  IN2P3-CNRS, Orsay, France}                                  
\centerline{$^{17}$LPNHE, IN2P3-CNRS, Universit\'es Paris VI and VII,         
                  Paris, France}                                              
\centerline{$^{18}$DAPNIA/Service de Physique des Particules, CEA, Saclay,    
                  France}                                                     
\centerline{$^{19}$IReS, IN2P3-CNRS, Universit\'e Louis Pasteur, Strasbourg,  
                France, and Universit\'e de Haute Alsace, Mulhouse, France}   
\centerline{$^{20}$Institut de Physique Nucl\'eaire de Lyon, IN2P3-CNRS,      
                   Universit\'e Claude Bernard, Villeurbanne, France}         
\centerline{$^{21}$III. Physikalisches Institut A, RWTH Aachen,               
                   Aachen, Germany}                                           
\centerline{$^{22}$Physikalisches Institut, Universit{\"a}t Bonn,             
                  Bonn, Germany}                                              
\centerline{$^{23}$Physikalisches Institut, Universit{\"a}t Freiburg,         
                  Freiburg, Germany}                                          
\centerline{$^{24}$Institut f{\"u}r Physik, Universit{\"a}t Mainz,            
                  Mainz, Germany}                                             
\centerline{$^{25}$Ludwig-Maximilians-Universit{\"a}t M{\"u}nchen,            
                   M{\"u}nchen, Germany}                                      
\centerline{$^{26}$Fachbereich Physik, University of Wuppertal,               
                   Wuppertal, Germany}                                        
\centerline{$^{27}$Panjab University, Chandigarh, India}                      
\centerline{$^{28}$Delhi University, Delhi, India}                            
\centerline{$^{29}$Tata Institute of Fundamental Research, Mumbai, India}     
\centerline{$^{30}$University College Dublin, Dublin, Ireland}                
\centerline{$^{31}$Korea Detector Laboratory, Korea University,               
                   Seoul, Korea}                                              
\centerline{$^{32}$CINVESTAV, Mexico City, Mexico}                            
\centerline{$^{33}$FOM-Institute NIKHEF and University of                     
                  Amsterdam/NIKHEF, Amsterdam, The Netherlands}               
\centerline{$^{34}$Radboud University Nijmegen/NIKHEF, Nijmegen, The          
                  Netherlands}                                                
\centerline{$^{35}$Joint Institute for Nuclear Research, Dubna, Russia}       
\centerline{$^{36}$Institute for Theoretical and Experimental Physics,        
                  Moscow, Russia}                                             
\centerline{$^{37}$Moscow State University, Moscow, Russia}                   
\centerline{$^{38}$Institute for High Energy Physics, Protvino, Russia}       
\centerline{$^{39}$Petersburg Nuclear Physics Institute,                      
                   St. Petersburg, Russia}                                    
\centerline{$^{40}$Lund University, Lund, Sweden, Royal Institute of          
                   Technology and Stockholm University, Stockholm,            
                   Sweden, and}                                               
\centerline{Uppsala University, Uppsala, Sweden}                              
\centerline{$^{41}$Lancaster University, Lancaster, United Kingdom}           
\centerline{$^{42}$Imperial College, London, United Kingdom}                  
\centerline{$^{43}$University of Manchester, Manchester, United Kingdom}      
\centerline{$^{44}$University of Arizona, Tucson, Arizona 85721, USA}         
\centerline{$^{45}$Lawrence Berkeley National Laboratory and University of    
                  California, Berkeley, California 94720, USA}                
\centerline{$^{46}$California State University, Fresno, California 93740, USA}
\centerline{$^{47}$University of California, Riverside, California 92521, USA}
\centerline{$^{48}$Florida State University, Tallahassee, Florida 32306, USA} 
\centerline{$^{49}$Fermi National Accelerator Laboratory, Batavia,            
                   Illinois 60510, USA}                                       
\centerline{$^{50}$University of Illinois at Chicago, Chicago,                
                   Illinois 60607, USA}                                       
\centerline{$^{51}$Northern Illinois University, DeKalb, Illinois 60115, USA} 
\centerline{$^{52}$Northwestern University, Evanston, Illinois 60208, USA}    
\centerline{$^{53}$Indiana University, Bloomington, Indiana 47405, USA}       
\centerline{$^{54}$University of Notre Dame, Notre Dame, Indiana 46556, USA}  
\centerline{$^{55}$Iowa State University, Ames, Iowa 50011, USA}              
\centerline{$^{56}$University of Kansas, Lawrence, Kansas 66045, USA}         
\centerline{$^{57}$Kansas State University, Manhattan, Kansas 66506, USA}     
\centerline{$^{58}$Louisiana Tech University, Ruston, Louisiana 71272, USA}   
\centerline{$^{59}$University of Maryland, College Park, Maryland 20742, USA} 
\centerline{$^{60}$Boston University, Boston, Massachusetts 02215, USA}       
\centerline{$^{61}$Northeastern University, Boston, Massachusetts 02115, USA} 
\centerline{$^{62}$University of Michigan, Ann Arbor, Michigan 48109, USA}    
\centerline{$^{63}$Michigan State University, East Lansing, Michigan 48824,   
                   USA}                                                       
\centerline{$^{64}$University of Mississippi, University, Mississippi 38677,  
                   USA}                                                       
\centerline{$^{65}$University of Nebraska, Lincoln, Nebraska 68588, USA}      
\centerline{$^{66}$Princeton University, Princeton, New Jersey 08544, USA}    
\centerline{$^{67}$Columbia University, New York, New York 10027, USA}        
\centerline{$^{68}$University of Rochester, Rochester, New York 14627, USA}   
\centerline{$^{69}$State University of New York, Stony Brook,                 
                   New York 11794, USA}                                       
\centerline{$^{70}$Brookhaven National Laboratory, Upton, New York 11973, USA}
\centerline{$^{71}$Langston University, Langston, Oklahoma 73050, USA}        
\centerline{$^{72}$University of Oklahoma, Norman, Oklahoma 73019, USA}       
\centerline{$^{73}$Brown University, Providence, Rhode Island 02912, USA}     
\centerline{$^{74}$University of Texas, Arlington, Texas 76019, USA}          
\centerline{$^{75}$Southern Methodist University, Dallas, Texas 75275, USA}   
\centerline{$^{76}$Rice University, Houston, Texas 77005, USA}                
\centerline{$^{77}$University of Virginia, Charlottesville, Virginia 22901,   
                   USA}                                                       
\centerline{$^{78}$University of Washington, Seattle, Washington 98195, USA}  
}                                                                             

\date{April 9, 2005}

\begin{abstract}
We have performed a search for neutral Higgs bosons produced in association
with bottom quarks in $p\bar{p}$ collisions,
using 260~\ipb\ of data collected with the \dzero\ detector in Run
II of the Fermilab Tevatron Collider. The cross sections for these
processes are enhanced in many extensions of the standard model
(SM), such as in its minimal supersymmetric extension at large
\tanb. The results of our analysis agree with expectations from the
SM, and we use our measurements to set upper limits on the
production of neutral Higgs bosons in the mass range of 90 to
150~\gev.
\end{abstract}


\pacs{12.38.Qk, 12.60.Fr, 13.85.Rm, 14.80.Cp} \maketitle

In two-Higgs-doublet models of electroweak symmetry breaking, such
as the minimal supersymmetric extension of the standard model
(MSSM)~\cite{2HDM}, there are five physical Higgs bosons:
two neutral $CP$-even scalars, \hboson\ and
\Hboson, with \Hboson\ being the heavier state; a neutral $CP$-odd state,
\Aboson; and two charged states, \Hboson$^\pm$. The ratio of the vacuum
expectation values of the two Higgs fields is defined as \tanb\ =
$v_2/v_1$, where $v_2$ and $v_1$ refer to the fields that couple to
the up-type and down-type fermions, respectively.
At tree level, the coupling of the $A$ boson to down-type quarks,
such as the \btag\ quark,
is enhanced by a factor of \tanb\ relative to the standard model (SM),
and the production
cross section is therefore enhanced by $\tan^{2}\beta$~\cite{MSSM_Higgs}.
At large \tanb, this is
also true either for the $h$ or $H$ boson depending on their mass.

For several representative scenarios of the MSSM, LEP experiments
have excluded at the 95\% C.L. a light Higgs boson with mass
$m_h$ \lt\ 92.9~\gev~\cite{leplimit}. At hadron colliders, neutral
Higgs bosons can be produced in association with \btag\ quarks,
leading to final states containing three or four \btag\ jets. The CDF
experiment at the Tevatron Collider performed a search
for these events in data from Run~I~\cite{cdfrunIlimit}.

\begin{figure}\centering
\includegraphics[width=3.0in]{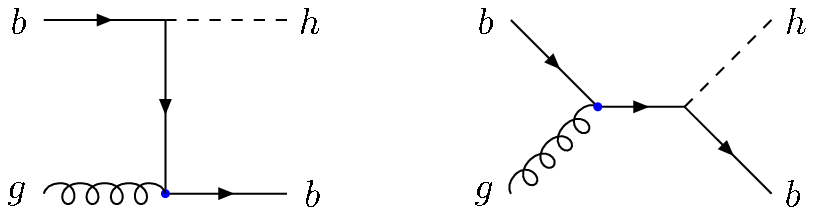}
\includegraphics[width=3.0in]{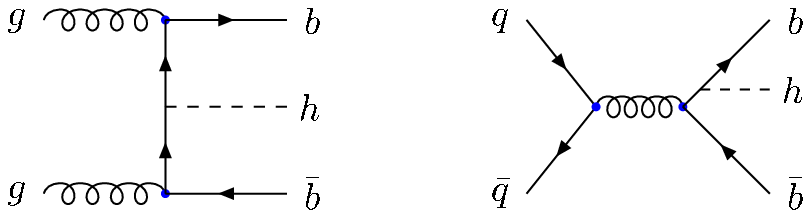}
\caption{Leading-order Feynman diagrams for 
neutral Higgs boson production in the five-flavor scheme (top) and
four-flavor scheme (bottom). } \label{fig:Feynman}
\end{figure}

Higgs boson production in association with \btag\ quarks in \ppbar\
collisions can be calculated in two ways: in the five-flavor
scheme~\cite{5fns}, only one \btag\ quark has to be present, while
in the four-flavor scheme~\cite{4fns}, two \btag\ quarks are
explicitly required in the final state. Both calculations are now
available at next-to-leading order (NLO), and agree within their
respective theoretical uncertainties~\cite{bh_NLO,top_NLO}.
Figure~\ref{fig:Feynman} illustrates these processes for \hboson\
production at leading order (LO), and analogous diagrams can be
drawn for the \Hboson\ and \Aboson\ bosons.

In this Letter, we assume $CP$-conservation in the Higgs sector. The
masses, widths, and branching fractions for the neutral Higgs bosons
into \bbbar\ pairs are calculated using the {\sc CPsuperH}
program~\cite{cpsuperh,carena}.
The current analysis is sensitive to \tanb\ in the range 50 -- 100,
and depends on the Higgs boson mass.
In this region of \tanb, the \Aboson\ boson is nearly degenerate in
mass with either the \hboson\ or the \Hboson\ boson, and their
widths are small compared to the di-jet mass resolution.
Consequently, we cannot distinguish between the \hboson/\Hboson\
and the \Aboson, and the total cross section for signal is
assumed to be twice that of the
\Aboson\ boson.
In the region of $m_A$ from 100 to 130~\gev, all three neutral Higgs
bosons can be degenerate in mass and produced
simultaneously~\cite{boos}. Nevertheless, the total cross section
still remains twice that of the \Aboson\ boson.
Using data collected by the \dzero\ detector from November 2002 to
June 2004, corresponding to an integrated luminosity of about
260~\ipb, we search for an excess in the invariant
mass distribution of the two leading transverse momentum (\pt) jets
in events containing three or more \btag\ quark candidates.

The D\O\ detector has a magnetic central tracking system surrounded
by a uranium/liquid-argon calorimeter, contained within a muon
spectrometer. The tracking system consists of a silicon microstrip
tracker (SMT) and a central fiber tracker (CFT), both located within
a 2~T solenoidal magnet~\cite{run2det}. The SMT and CFT have designs
optimized for tracking and vertexing at pseudorapidities
$|\eta|<2.5$, where $\eta = -\ln(\tan(\theta/2))$ and $\theta$ is
the polar angle with respect to the proton beam direction ($z$). The
calorimeter has a central section (CC) covering up to $|\eta|
\approx 1.1$, and two end calorimeters (EC) extending coverage to
$|\eta|\approx 4.2$, all housed in separate
cryostats~\cite{run1det}. The calorimeter is divided into an
electromagnetic part followed by fine and coarse hadronic sections.
Scintillators between the CC and EC cryostats provide additional
sampling of developing showers for $1.1<|\eta|<1.4$. The muon system
consists of a layer of tracking detectors and scintillation trigger
counters in front of 1.8~T toroidal magnets, followed by two similar
layers behind the toroids, which provide muon tracking for
$|\eta|<2$. The luminosity is measured using scintillator arrays
located in front of the EC cryostats, covering $2.7<|\eta|<4.4$. The
trigger system comprises three levels (L1, L2, and L3),
each performing an increasingly detailed event reconstruction in
order to select the events of interest.

The large cross section for multijet production necessitates a
specialized trigger to maximize signal acceptance while providing
reasonable rates. This trigger at L1 requires signals in at least
three calorimeter towers of size
$\Delta\eta\times\Delta\varphi=0.2\times0.2$ (where $\varphi$ is the
azimuthal angle), each with transverse energy $E_T > 5$~GeV; three
clusters and $H_T^{L2}>50$~\gev\ at L2 ($H_T^{L2} \equiv$ scalar sum
of the L2 clusters \et\ with \et\gt 5 \gev), and three jets with
$\pt>15$~\gev\ at L3. A total of 87 million events were selected
off-line with one jet of $\pt>20$~\gev\ and at least two more jets
with $\pt>15$~\gev.
Jets are reconstructed using a Run~II cone
algorithm~\cite{RunIIcone} with radius
$\Delta{\cal{R}}=\sqrt{(\Delta \eta)^2+(\Delta \varphi)^2}<0.5$, and are
then required to pass a set of quality criteria.
To be accepted for further analysis, jets with
$\pt>15$~\gev\ must have $|\eta|<2.5$. The jet energies are
corrected to the particle level
using $\eta$-dependent scale factors. Events with up to five jets
are selected if they have a primary vertex position
$|z|<35~\mathrm{cm}$ and at least three jets with corrected
$\pt>35$, 20, and 15~\gev. Depending on the hypothesized Higgs boson
mass, the final selections are chosen to optimize the expected
signal significance, defined as $S/\sqrt{B}$, where $S$ ($B$) refers
to the number of signal (background) events.
Jets containing \btag\ quarks are identified using a secondary
vertex (SV) tagging algorithm. A jet is tagged as a \btag-jet if it
has at least one SV within $\Delta{\cal{R}}<0.5$ of the jet axis
and a transverse displacement from the primary vertex that
exceeds five times the displacement uncertainty. Jets are \btag\ tagged up to
$|\eta|<2.5$, although the \btag\ tagging is about twice as
efficient in the central region ($|\eta|<1.1$) because of the CFT
coverage.
The \btag\ tagging efficiency is $\approx$~55\% for central
\btag-jets of $\pt>35$~\gev, with a light quark (or gluon) tag rate
of about 1\%.

Signal events were simulated using the
{\sc pythia}~\cite{pythia} event generator followed by the full
\dzero\ detector simulation and reconstruction chain. {\sc pythia}
minimum-bias events were added to all generated events, using a
Poisson probability with a mean of 0.4 events to match the
instantaneous luminosities at which the data were taken ($1-6\times
10^{31} {\mathrm{cm}}^{-2}\mathrm{s}^{-1}$). The \bh\ events, with
\hboson\rarrow\bbbar, were generated for Higgs boson masses from 90
to 150 \gev. Reconstructed jets in simulated events were corrected
to match the jet reconstruction and identification efficiencies in
data. The energy of simulated jets was smeared to match the measured
jet energy resolution. The \pt\ and rapidity spectra of the Higgs
bosons from {\sc pythia} were compared to those from the NLO
calculation~\cite{5fns}. The shapes were similar, indicating that
the {\sc pythia} kinematics are approximately correct.
The simulated events were weighted to match the \pt\
spectrum of the Higgs boson given by NLO, resulting in
a 10\% reduction of the overall
signal efficiency.

Of all SM processes, multijet production is the major source of
background. This background is determined from data by normalizing
distributions outside of the signal region. As a cross-check, we
also compare data with simulations. {\sc alpgen}~\cite{alpgen} is
used to generate three samples of events for \bbj\ and \bbjj\ with
$j$ corresponding to up, down, strange or charm quarks, or gluons,
and \bbbar\bbbar\ final states with generator-level requirements:
$p_T^b>25$~\gev, $p_T^j>15$~\gev, $\aeta<3.0$, and
$\Delta{\cal{R}}>0.4$ between any two final-state partons. These
selections do not introduce significant bias because the final
sample contains much harder jets, after the application of trigger
and \btag-tagging requirements. Samples of \bbj\ and \bbjj\ are
added together, but the \bbjj\ sample is weighted by 0.85 to match
the jet multiplicity observed in doubly \btag-tagged data. The cross
sections obtained from {\sc alpgen} are 8.9~nb, 3.9~nb, and 60~pb,
for the respective three states. All other backgrounds are expected
to be small and are simulated with {\sc pythia}:
\ppbar\rarrow\zboson(\rarrow\bbbar)+jets, \ppbar\rarrow\zb, and
\ppbar\rarrow\ttbar. Cross sections of
1.2~nb, 40~pb~\cite{Zb_NLO},
and 7~pb are assumed, respectively.



There are two main categories of multijet background. One contains
genuine heavy-flavor (HF) jets, while the other has only light-quark
or gluon jets that are mistakenly tagged as \btag-quark jets, or
correspond to gluons that branch into nearly collinear \bbbar\
pairs. Using the selected data sample, before the application of
\btag-tagging requirements, the probability to \btag-tag a jet is
measured as a function of its \pt\ in three \aeta\ regions. These
functions are called ``mis-tag'' functions. They are corrected for
the contamination from true HF events
by subtracting the estimated fraction of \bb\ events in the
multijet data sample (1.2\%), obtained from an initial fit to the doubly
\btag-tagged data.
These corrected mis-tag functions are then used to estimate
the mis-tagged background, by applying them to every jet
reconstructed in the full data sample.

\begin{figure}\centering
\includegraphics[width=3in]{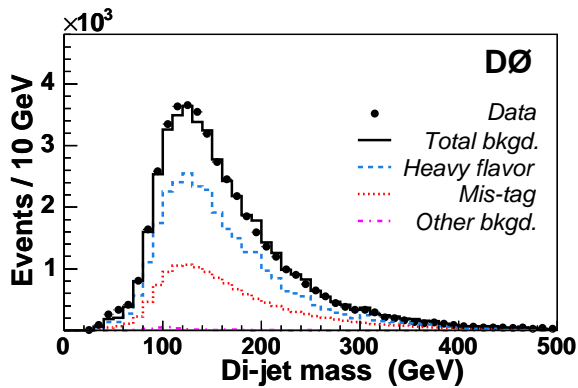}
\caption{Fit of the invariant mass spectrum of the two leading \pt\
jets in the doubly \btag-tagged data to a sum of backgrounds:
mis-tags derived from data (dotted), \bb\ (dashed), and other
backgrounds (\zboson(\rarrow\bbbar)+jets, \zb, \ttbar\ and
\bbbar\bbbar) (dashed-dotted).} \label{fig:Final_QCD_fit}
\end{figure}

In order to test the modeling of the mis-tag background, the high
statistics doubly \btag-tagged data is compared to simulations
first, before extrapolating to the triply \btag-tagged background.
The expected signal contribution to the doubly \btag-tagged data is
negligible. The comparison in invariant mass spectrum of the two
jets of highest \pt\ (not necessarily the two \btag-tagged jets) in
the doubly \btag-tagged data with the expected background is shown
in Fig.~\ref{fig:Final_QCD_fit}. The \btag-tagging in this analysis
does not distinguish between contributions from bottom and charm
events. However, the efficiency for tagging a $c$-jet is known from
simulations to be about $1/4$ of that for tagging a \btag-jet.
Therefore, when two \btag-tags are required, the fraction of \cc\
events relative to \bb\ events will be a factor of
$\approx$~16 lower after tagging.
We have estimated the fractions of \ccjj\ to \bbjj\ prior to
\btag-tagging using the {\sc madgraph} Monte Carlo
generator~\cite{madgraph}. The \ccjj\ cross section is 22\% higher
than \bbjj\ for the same generator-level selections. Therefore,
the contribution of \cc\ in the doubly \btag-tagged data sample is
expected to represent about
8\% of the events. Thus, when we refer to the \bb\ normalization, it
should be understood that approximately 8\% of the events are from
the \cc\ process. After these corrections for \cc\ events, the HF
multijet processes are only a factor of 1.08 higher in data than
predicted by {\sc alpgen}.
The shape of the estimated background agrees well with the data over
the entire invariant mass region.



\begin{figure}\centering
\includegraphics[width=3in]{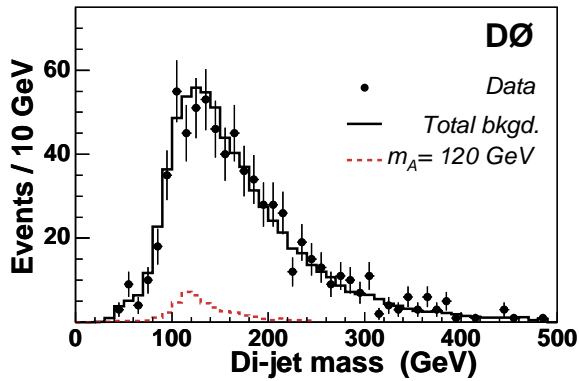}
\caption{Invariant mass spectrum of two leading jets in events with
at least three \btag-tagged jets, estimated background, and the
signal for a 120~\gev\ Higgs boson that can be excluded at the 95\%
C.L.} \label{fig:mh_limit}
\end{figure}

To estimate the background for triply \btag-tagged events, the
mis-tag function is applied to the non-\btag-tagged jets in the
doubly \btag-tagged events. This provides the shape of the multijet
background distribution with at least three \btag-tagged jets. This
neglects any contributions from processes with more than two true
\btag-jets, such as from \bbbar\bbbar\ and
\zboson(\rarrow\bbbar)\bbbar\ production. However, the shapes of
these backgrounds from simulations are similar to those of the
doubly \btag-tagged spectra, and their rates are small. The overall
background normalization is therefore determined by fitting the
leading two jets invariant mass spectrum in triply \btag-tagged
events outside of the hypothesized signal region
to the estimated shape for triply
\btag-tagged background. The systematic effect on the
normalization of the background from any signal contributing
outside the search window was studied and found to be small relative
to other uncertainties, as described below.


The selections in this analysis can be grouped into trigger level,
kinematic (\pt, \peta, \nj), where \nj\ is the number of untagged
jets, and \btag-tagging. Table~\ref{table:Acceptance_Cutflow} shows
the acceptances for each set of criteria made in the analysis, for
six values of Higgs boson mass. The systematic uncertainty on signal
acceptance is nearly independent of assumed $m_A$, and is dominated
by the uncertainty on \btag-tagging efficiency ($\pm$15\%), followed
by uncertainties on jet energy scale, resolution and identification
efficiency ($\pm$9\% in sum). These uncertainties are calculated by
repeating the analysis with each value changed by $\pm$ one standard
deviation (sd). The systematic uncertainties corresponding to
uncertainties in \pt\ distributions for simulated signal at NLO, the
integrated luminosity, and the trigger efficiency are
found to be $\pm$5\%, $\pm$6.5\%, and $\pm$9\%, respectively. These
uncertainties, added in quadrature, result in a total systematic
uncertainty of $\pm$21\%.

The accuracy in modeling the shape of the background distribution can be
estimated from the $\chi^{2}/dof$ between the estimated
background and the data. The statistical error associated with
the uncertainty in the normalization of the background (from the fit
outside the signal region) is multiplied by
$\sqrt{\chi^{2}/dof}$. The background uncertainty is estimated to
be $\lsim$ 3\%. The systematic uncertainty arising from the width chosen
for the search window
is evaluated by varying
it from less than the resolution to $\pm 1.8$ sd, centered on the peak value.
The resulting change in background normalization is much smaller
than from other sources of background uncertainties.



\begin{table} \centering
\caption{Signal acceptances for each set of criteria (in
\%).}
\begin{ruledtabular}
\begin{tabular}{ccccc}
$m_A$ (\gev) & Trigger & Kinematic & \btag-tag & Total \\ \hline
90  & 44 & 18 & 3.5 & 0.3 \\
100 & 45 & 24 & 3.5 & 0.4 \\
110 & 56 & 24 & 3.9 & 0.5 \\
120 & 60 & 27 & 4.2 & 0.7 \\
130 & 65 & 29 & 4.3 & 0.8 \\
150 & 76 & 31 & 4.4 & 1.0 \\
\end{tabular}
\end{ruledtabular}
\label{table:Acceptance_Cutflow}
\end{table}


\begin{figure}\centering
\includegraphics[width=3in]{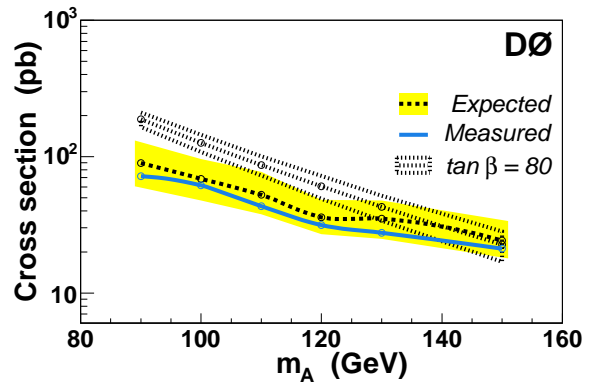}
\caption{The expected and measured 95\% C.L. upper limits on the signal
cross section as a function of $m_A$. The band
indicates the $\pm 1$ sd range on the expected limit. Also shown
is the cross section for the signal at \tanb\ = 80 in the ``no
mixing'' scenario of the MSSM, with the theoretical uncertainty
indicated by the overlaid band.} \label{fig:mA_cs}
\end{figure}

A modified frequentist method is used to set limits on the
production of signal~\cite{mclimit}. The di-jet invariant mass
distributions in triply \btag-tagged events of data, simulated
signal, and the normalized background were used as inputs. The value
of \tanb\ was varied
until the confidence level for signal ($CL_S$) was $<5$\%.
Figure~\ref{fig:mh_limit} shows the data, background, and simulated
signal at the exclusion limit, for $m_A$ = 120~\gev. This is
converted to a cross section limit for signal production in
Fig.~\ref{fig:mA_cs}, which also shows the expected MSSM Higgs boson
production cross section as a function of $m_A$ for \tanb\ = 80, and
the median expected limit with the background-only hypothesis along
with its $\pm$ 1 sd range. The NLO cross sections and their
uncertainties from parton distribution functions (PDF) and scale
dependence are taken from Refs.~\cite{5fns,top_NLO}. The MSSM cross
section shown in Fig.~\ref{fig:mA_cs} corresponds to no mixing in
the scalar top quark sector~\cite{tev}, or $X_{t}$ = 0, where
${X_{t} = A_{t} - \mu\cot\beta}$, $A_{t}$ is the tri-linear
coupling, and the Higgsino mass parameter $\mu = -0.2$~TeV. We also
interpret our results in the ``maximal mixing'' scenario with $X_{t}
= \sqrt6 \times M_{\text{SUSY}}$, where $M_{\text{SUSY}}$ is the
mass scale of supersymmetric particles, taken to be 1~TeV.

\begin{figure}\centering
\includegraphics[width=3in]{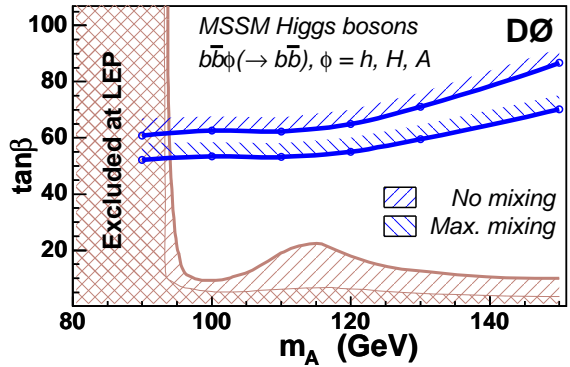}
\caption{The 95\% C.L. upper limit on \tanb\ as a function of $m_A$
for two scenarios of the MSSM, ``no mixing'' and ``maximal mixing.''
Also shown are the limits obtained by the LEP
experiments for the same two scenarios of the MSSM~\cite{leplimit}.}
\label{fig:mA_tanb}
\end{figure}

Results for both scenarios of the MSSM are shown in
Fig.~\ref{fig:mA_tanb} as limits in the \tanb\ versus $m_A$ plane.
The present D\O\ analysis, based on 260~\ipb\ of data, excludes a
significant portion of the parameter space, down to \tanb = 50,
depending on $m_A$ and the MSSM scenario assumed.


We thank the authors of Refs.~\cite{5fns,top_NLO,tev} for valuable
discussions. 
We thank the staffs at Fermilab and collaborating institutions, 
and acknowledge support from the 
DOE and NSF (USA),
CEA and CNRS/IN2P3 (France),
FASI, Rosatom and RFBR (Russia),
CAPES, CNPq, FAPERJ, FAPESP and FUNDUNESP (Brazil),
DAE and DST (India),
Colciencias (Colombia),
CONACyT (Mexico),
KRF (Korea),
CONICET and UBACyT (Argentina),
FOM (The Netherlands),
PPARC (United Kingdom),
MSMT (Czech Republic),
CRC Program, CFI, NSERC and WestGrid Project (Canada),
BMBF and DFG (Germany),
SFI (Ireland),
A.P.~Sloan Foundation,
Research Corporation,
Texas Advanced Research Program,
Alexander von Humboldt Foundation,
and the Marie Curie Fellowships.
%

\end{document}